\documentclass[
aps,%
11pt,%
final,%
notitlepage,%
oneside,%
onecolumn,%
nobibnotes,%
nofootinbib,%
superscriptaddress,%
noshowpacs,%
centertags]%
{revtex4}

\begin{document}
\selectlanguage{english}

\title{
Relative distances and peculiar velicities of 140 groups and
clusters of galaxies at low redshifts: the Hubble diagram
}
\author{\firstname{F.G.}~\surname{Kopylova,}}
\affiliation{\saoname}

\author{\firstname{A.I.}~\surname{Kopylov}}
\affiliation{\saoname}


\begin{abstract}
\noindent

To determine the relative distances and peculiar velocities of 140 groups
and clusters of galaxies at low redshifts ($z$ < 0.12), we used the
fundamental plane (FP) of early-type galaxies (from the Sloan Digital
Sky Survey (SDSS) data). We constructed the Hubble diagram for the
relative distances of galaxy groups/clusters versus their radial velocities
in the cosmic microwave background (CMB) reference frame in
the flat $\Lambda$ cold dark matter ($\Lambda$CDM) model
($\Omega_m=0.3$, $H_0=70$~km~s$^{-1}$ Mpc$^{-1}$). We have found that
the standard logarithmic deviation for groups and clusters of galaxies on
the Hubble diagram (minus peculiar velocities) is $\pm0.0173$ ($N$ = 140),
which corresponds to a deviation of $70\pm2.8$~km~s$^{-1}$ Mpc$^{-1}$ in the
Hubble  constant. For a sample of galaxy systems ($N$ = 63), the X-ray
luminosity of which is in an interval of (0.151--4)~$\times 10^{44}$~erg/s,
this quantity turned out to be $70\pm2.1$~km~s$^{-1}$ Mpc$^{-1}$.
The root-mean-square deviations of peculiar velocities with quadratic
accounting for errors are $<V_{pec}^2>^{1/2}$ = $714\pm7$~km/s and
$600\pm7$~km/s, respectively. For five large superclusters of
galaxies from the SDSS region,
the average peculiar velocity relaive to the CMB reference frame is
$+240\pm250$~km/s. We detected no outflow of galaxy systems from the
void
(Giant Void, $\alpha \approx 13^h, \delta \approx 40^\circ, z \approx 0.107$)
formed by groups and clusters of galaxies.

\vspace{0.3cm} \noindent

\end{abstract}

\maketitle

\section{INTRIDUCTION}

A large-scale structure of the Universe is arranged
in a cellular way. Its main elements are halos of dark
matter---galaxies, groups and clusters of galaxies
concentrated in filaments enveloping low-density regions, i.e., voids.
The first studies of the large-scale
structure elements were performed in papers \cite{Gregory, Joeveer,
Kirshner, de Lapparent, Kopylov88, Bond, Einasto01}.
The authors of \citep{Zeldovich} proposed the scenarios for explaining
the formation and evolution of the large-scale
structure and note that the latter is mostly represented
by empty regions---voids. Voids in the distribution of
galaxy clusters were studied, for example, in papers \cite{Batuski,
Tully, Stavrev, Kopylov02, Einasto11}.

The gravitational interaction of the large-scale
structure elements is a main cause of peculiar velocities
of galaxies and galaxy clusters. The peculiar velocity
of galaxy clusters at small $z$ can be estimated as
follows
\begin{equation}
V_p \approx cz_{obs} - cz_H \approx cz_{obs} -H_oD,
\end{equation}
where $D$ is the comoving distance of a galaxy $H_o$ is the Hubble constant.

To determine the peculiar velocities of galaxy clusters
relative to the Hubble flow, one should measure
the relative distances of galaxy systems with one of the
techniques sensitive to the distance. The concept of
the fundamental plane (FP) of early-type galaxies
\cite{Dressler, Djorgovski} is widely used to study the
properties of these galaxies and to determine the relative distances
and peculiar velocities of galaxy clusters (for example, \cite{Wegner, Hudson,
da Costa, Batiste}).
The FP is an empirical relationship between the central velocity dispersion
of stars
in  the galaxy $\sigma$, the physical effective radius $R_e$, and the average
surface brightness $\mu_e$ within the effective radius.

Earlier, we already used the FP for a large sample of systems of
galaxies to determine the relative distances and peculiar velocities of
galaxy clusters in the Leo, Hercules (Her), Bootes (Boo), Coropa Borealis
(CrB) superclusters
on the basis of the Sloan Digital Sky Survey (SDSS) catalog (Data Release
(DR) 8; \cite{Aihara}) \cite{Kopylova14, Kopylova17, Kopylova21}.
SDSS (DR8 catalog data \cite{Aihara}) \cite{Kopylova14, Kopylova17,
Kopylova21}. In a paper \cite{Kopylova07}, we reported
the results of analogous measurements for the Ursa
Major (UMa) supercluster based on the SDSS DR4 data.

In the DR8 catalog, the previous errors in the
image processing, especially for large galaxies, were
taken into account. In the above mentioned papers \cite{Kopylova17,
Kopylova21}, to measure the observed relative distances of the systems
of galaxies, we used the FP, which had
been already obtained in \cite{Saulder}. In that paper, the evolution
in the luminosity of early-type galaxies was accounted for and the
evolutionary parameter $Q=1\,.\!\!^{\rm m}07z$ was derived.

As is known, in the expanding Universe, the surface
brightness of an object changes as $SB \propto (1+z)^{-4}$ ($z$ is the
redshift of the object, and SB is its surface brightness),
which describes the  cosmological dimming effect in the surface brightness.
The factor $(1+z_H)^{-2}$ appears due to the Universe expansion,
while the factor $(1+z_{obs})^{-2}$, due to relativistic effects
caused by the radial proper motion. Hence, the
logarithmic correction for the surface brightness dimming of galaxies in
the expanding Universe (see, e.g. \cite{Mohr}) can be written in the
following way
\begin{equation}
C = 5\,\log(1+z_{obs})+5\,\log(1+z_H).
\end{equation}

Moreover, in a paper \cite{Saulder}, the evolution of the average surface
brightness of early-type galaxies with redshift (the evolutionary parameter is
$Q_r$ = 2.2z (expressed in $mag/arcsec^2$)) was determined with
the cosmological dimming of the surface brightness with changing $z$ as
$10\,\log(1+z_{CMB}$) was accounted for, and the
evolution of the stellar magnitudes of galaxies was ignored.

In mathematical terms, the accounting for the both corrections, $Q$ and
$Q_r$, in SB is the same.
In this paper, we show that, if only the first part of the surface brightness
correction caused by the motion of galaxies
$5\,\log(1+z_{obs}$) is taken into account, the evolution of
the average surface brightness with increasing $z$ is
$Q_r$ = 3.76~(expressed in $mag/arcsec^2$).

In addition to superclusters of galaxies, our present sample includes
the groups and clusters of galaxies located in a region  of the Giant Void
(GV) in the distribution of galaxy clusters
($\alpha \approx 13^h, \delta \approx 40^\circ, z \approx 0.107$).
The GV diameter is the maximumal diameter of a sphere containing no galaxy
clusters $R \geq$ 1, and it amounts to 214~Mpc.
From our earlier observations with the 6- and 1-m telescopes at the Special
Astrophysical Observatory (SAO) of the Russian Academy of Sciences, which
covered 17 clusters of galaxies and were based on the Kormendy relation, we
obtained that there is no outflow of galaxy systems caused by the
mass deficiency in the void \cite{Kopylov02}.

In the present paper, we redefine the relative distances and peculiar
velocities of galaxy clusters around the GV with the use of the other
method---the method
based on the FP of early-type galaxies. We will consider the entire sample
of similarly performed measurements of the relative distances for 140 groups
and clusters of galaxies as a whole. One of the main purposes of the study
is to verify the standard cosmological model by the distances and peculiar
velocities in a large sampling of galaxy systems (the Hubble diagram). For
this, we drew the data from the SDSS (DR7 and DR8) and from the NASA/IPAC
Extragalactic Database (NED).

The paper is organized in the following way. In Section 2, we describe
the stages of building the FP --- how we selected early-type galaxies for
the sample---and present the common FP. In Section 3, the relative
distances for the groups/clusters of galaxies are determined. In Section 4,
we calculate the peculiar velocities of the groups/clusters of galaxies
around the void and the peculiar velocities of superclusters of galaxies.
In Section 5, the Hubble diagram for the entire sample is presented and the
deviations from it are estimated. Finally, the results are listed. For
this analysis, we used the standard $\Lambda$CDM cosmology with the following
parameters:
$H_0=70$~km~s$^{-1}$ Mpc$^{-1}$, $\Omega_m=0.3$, $\Omega_{\Lambda}=0.7$.

\section{THE FUNDAMENTAL PLANE OF EARLY-TYPE GALAXIES}
\subsection{Description of Sampling}

In total, our sample covers 140 groups and clusters,
each of which contains more than three early-type galaxies within the
chosen radius $R_{200}$. The sample parameters are the following:
0.020 < $z_{CMB}$ < 0.120 (in addition, there are three clusters with
$z_{CMB}$ > 0.120) and  200 < $\sigma$ < 1104~km/s.
The most distant systems of galaxies considered here are in the GV region,
which we studied earlier. In a paper \cite{Kopylov02}, we selected 17
clusters of galaxies. However, according to the SDSS data, three clusters
(A\,1298, A\,1700, A\,1739) contain few galaxies described by spectral
information. Consequently, we do not consider them here, but add five more
clusters of galaxies located in this region. The sample for the region
around the Giant Void totally contains
19 groups and clusters of galaxies,  the redshifts of which are
0.07 < $z_{CMB}$ < 0.15.

The dynamical characteristics of galaxy systems are estimated on the basis
of measurements of the heliocentric radial velocity and the
one-dimensional dispersion of radial velocities, from which the virial mass
within the empirical radius $R_{200}$ is calculated under the assumption that
$M(r)\propto r$.
The radius $R_{200}$ is close to the virial one, and within
its limits the density of the galaxy systems 200 times exceeds the critical
density of the Universe. The radius $R_{200}$ can be estimated by the
formula $R_{200} = \sqrt {3} \sigma /10H(z)$ Mpc \cite{Carlberg}.
Under the assumption $M_{200} \simeq M_{vir}$, the mass within $R_{200}$ is
$M_{200} = 3G^{-1}R_{200}\sigma_{200}^{2}$ . The measurements of the
parameters of galaxy clusters are described at length, for example,
in a paper \cite{Kopylova17}.

\subsection{Selection of Early-Type Galaxies}

Early-type galaxies within the $R_{200}$ radius was
selected over all of the galaxiy clusters in the same way.
As in a paper \cite{Kopylova17}, we applied
the following criteria to the parameters of galaxies:

---the central dispersion in the velocities of stars is
100<$\sigma$<420 km/s;

---the parameter, characterizing the contribution of the de
Vaucouleurs profile into the surface brightness profile, is
$\rm fracDeV_r\geq 0.8$;

---the concentration index, which is the ratio of the radii
containing 90 and 50\% of the Petrosian fluxes, is $r_{90}/r_{50}\geq2.6$;

---the restriction by color is $\Delta (u-r)>-0.2$ in order to
exclude late-type  galaxies, where\linebreak \mbox{$\Delta (u-r)$ = $(u-r)+0.108\,M_r-0.3$}
\cite{Kopylova13};

---the axes ratio of galaxies is \mbox{$\rm deVAB\geq 0.3$};

---the signal-to-noise ratio in the spectra of galaxies is  $\rm snMedian>10$;

---the limiting stellar magnitude for our sample corresponds to the
spectral limit of the SDSS, which is equal to the Petrosyan stellar magnitude
(i.e., not uncorrected for absorption),
and amounts to $17\,.\!\!^{\rm m}77$ in the $r$  filter \cite{Strauss}.

The number of considered galaxies is of key importance for determining the
relative  distances of galaxy clusters, since the standard error in the
average distance is equal to the standard deviation divided by $\sqrt(N)$.
From the SDSS catalog, we took the parameters of galaxies
obtained by fitting the observed profile of galaxies by the de Vaucouleur
profile.
All of the corrections, which were introduced according to a paper \cite{Saulder},
are the following: (1) the aperture correction for dispersion in the
radial velocities
$\sigma_0=\sigma_{sdss}(r_{fiber}/(r_{cor}/8))^{0.04}$ (here,
$r_{cor}=r_{dev}\,\sqrt (b/a)$ is the radius of galaxy with accounting for
its
ellipticity, and $r_{dev}$ is the model radius of an early-type galaxy);
(2) the correction for absorption in the Galaxy (SDSS data); and (3) the
$K$ correction \cite{Chilingarian}.
The radial velocities of galaxy clusters were reduced to the relict
background (CMB) frame, and the corrections were taken from the NED.

Usually, the average effective surface brightness is written as follows
\begin{equation}
<\mu_e> = m_{dev}+2.5\,\log(2\pi\,r_{cor}^2)-K(z)-10\,\log(1+z).
\end{equation}

As has been already noted, we divided the correction for cosmological dimming
of the SB, $10\,\log(1+z)$, into two components (see Eq. (2)) to account for
the relativistic effects and the change in the Universe geometry.
In Eq. (2), $z_{obs}$ is the measured redshif, which contains the peculiar
velocity of the object, and $z_{H}$ ($z_{FP}$ in our case) is the redshift
corresponding to the true cosmological distance determined from the FP.

\begin{figure*}[p]
\setcaptionmargin{5mm}
\onelinecaptionsfalse
\includegraphics[scale=0.63,angle=0]{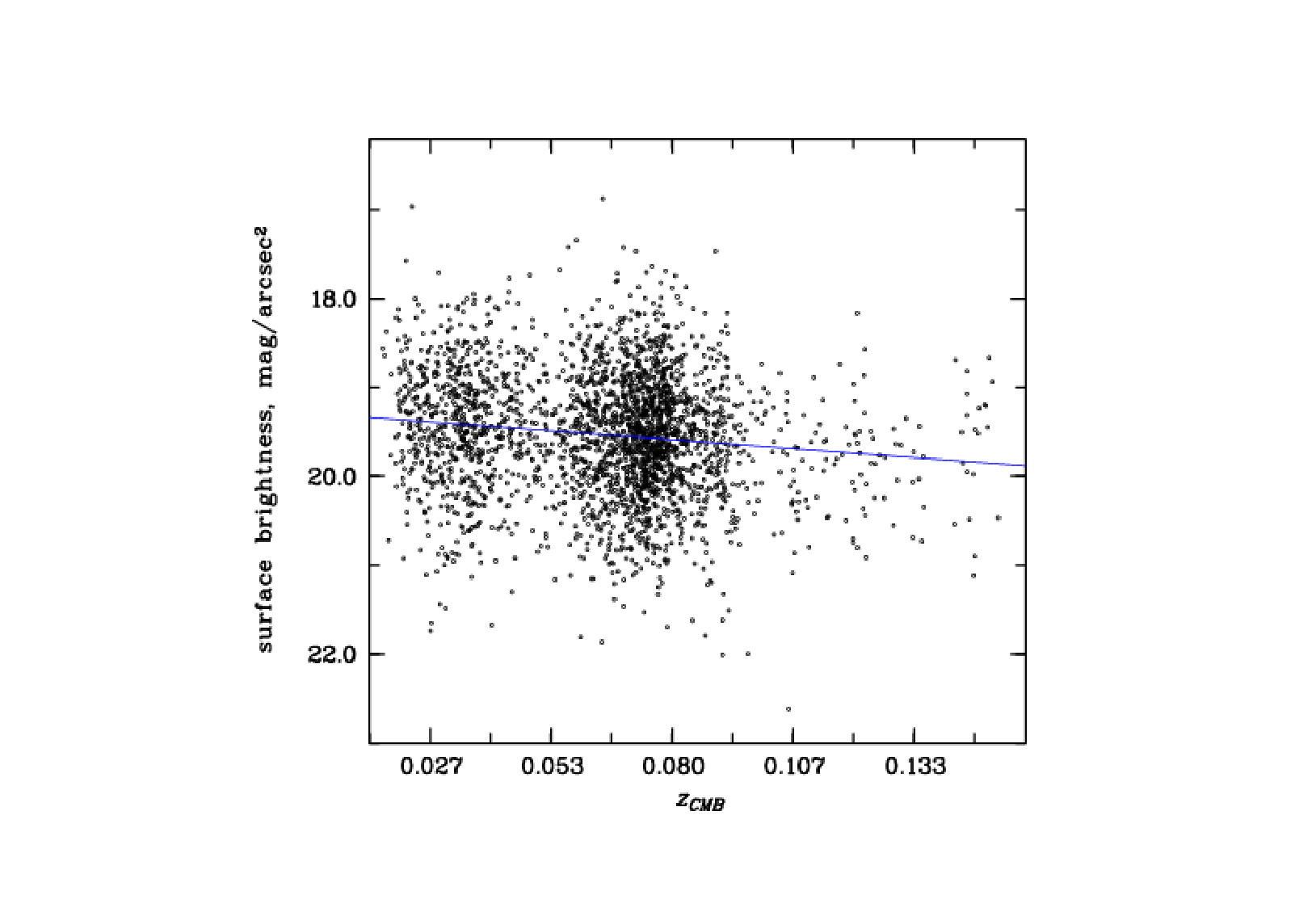}
\captionstyle{normal}
\caption{
The average surface brightness of early-type galaxies in a dependence
on $z_{CMB}$. The line corresponds to the regression relationship:
$<\mu_e> = 3.76(\pm0.56)z+19.285(\pm0.04)$.
}
\label{mue}
\end{figure*}

In the SB, we took into account only the first part of the
correction, while the second part is accounted for in the
zero point of the FP, as in \cite{Kopylova17},
when the Hubble dependence is determined (see the next subsection).
For our sample of early-type galaxies ($N$ = 2654), we used the first
part of the correction $5\,\log(1+z_{obs}$) to determine the dependence
$<\mu_e>$ on the redshift and derived
$<\mu_e> = 3.76(\pm0.56)z+19.285(\pm0.04)$.
In Fig.~\ref{mue}, we show the obtained dependence for the redshifts
range  $z_{CMB}$ ranging from 0.02 to 0.145.

\subsection{Determination of Distances by the FP}

For 2654 early-type galaxies, which we selected according to the above
criteria, the common FP was built in the comoving coordinate system with the
least-squares method. The equation for the FP has the following form:

\begin{eqnarray}
\log R_e(kpc)&=&(0.991\pm0.124)\,\log \sigma \nonumber \\
	     &+&(0.318\pm0.020)\,<\mu_e>+\gamma,
\end{eqnarray}
where $R_e$ is the effective radius of a galaxy expressed in kiloparsecs,
$<\mu_e>$ is the average effective surface brightness within this radius,
$\sigma$ is the dispersion of the radial velocities of the stars,
and $\gamma$ is the zero point of the FP that depends on the distances
of galaxies.
In our sample, a value of the zero point $\gamma$ = -$8.066\pm0.003$ was
obtained for the assumed standard $\Lambda$CDM cosmology.
The standard deviation of the FP zero point is 0.071, which is corresponds
to an error of $\sim$ 16\% in determining the distance of one galaxy.

The formal error in determining the distance of a cluster depends on the
number of considered galaxies and varies from 2\% to 12\%.
The zero point changes with the distance of galaxies if $\log R_e$ is measured
in arcseconds.
The residual deviations from the FP, $\Delta \gamma = \log R_e (arcsec)-
0.991\log \sigma-0.318<\mu_e>-\gamma$, do not depend on the central
dispersion  of stars in the galaxies.
We used this circumstance to specify more accurately the sample of the
previously selected galaxies in each of the clusters \cite{Kopylova17}.
We empirically found that almost all of the deviations in the zero points
$\gamma$ of galaxies from the mean zero point of the cluster does not
exceed 2$\sigma$.

The limiting stellar magnitude of our sample varies from system to system.
We accepted the same limit for all galaxy systems, $M_r = -21^m$,
determined the distances, and compared them to those obtained with
the use of individual limits. The differences between the distances are
within
$\pm5.7\%$, while the average deviation from this value is zero. Thus,
variations in the limiting magnitude of the galaxy systems produce
no substantial effect on the derived peculiar velocities and are almost
within the limits of their errors. This is especially true for distant
galaxy clusters of our sample, in which the peculiar velocities are very
unaccurately measured only by bright galaxies. Since our sample included
only ten galaxy systems containing less than seven galaxies, we ignored
the Malmquist bias in the distances. For the galaxy systems with more
than seven members, a homogeneous effect of the Malmquist bias is less than
1.5\% (\cite{Jorgensen}.

\begin{figure*}[p]
\setcaptionmargin{5mm}
\onelinecaptionsfalse
\includegraphics[scale=0.63,angle=0]{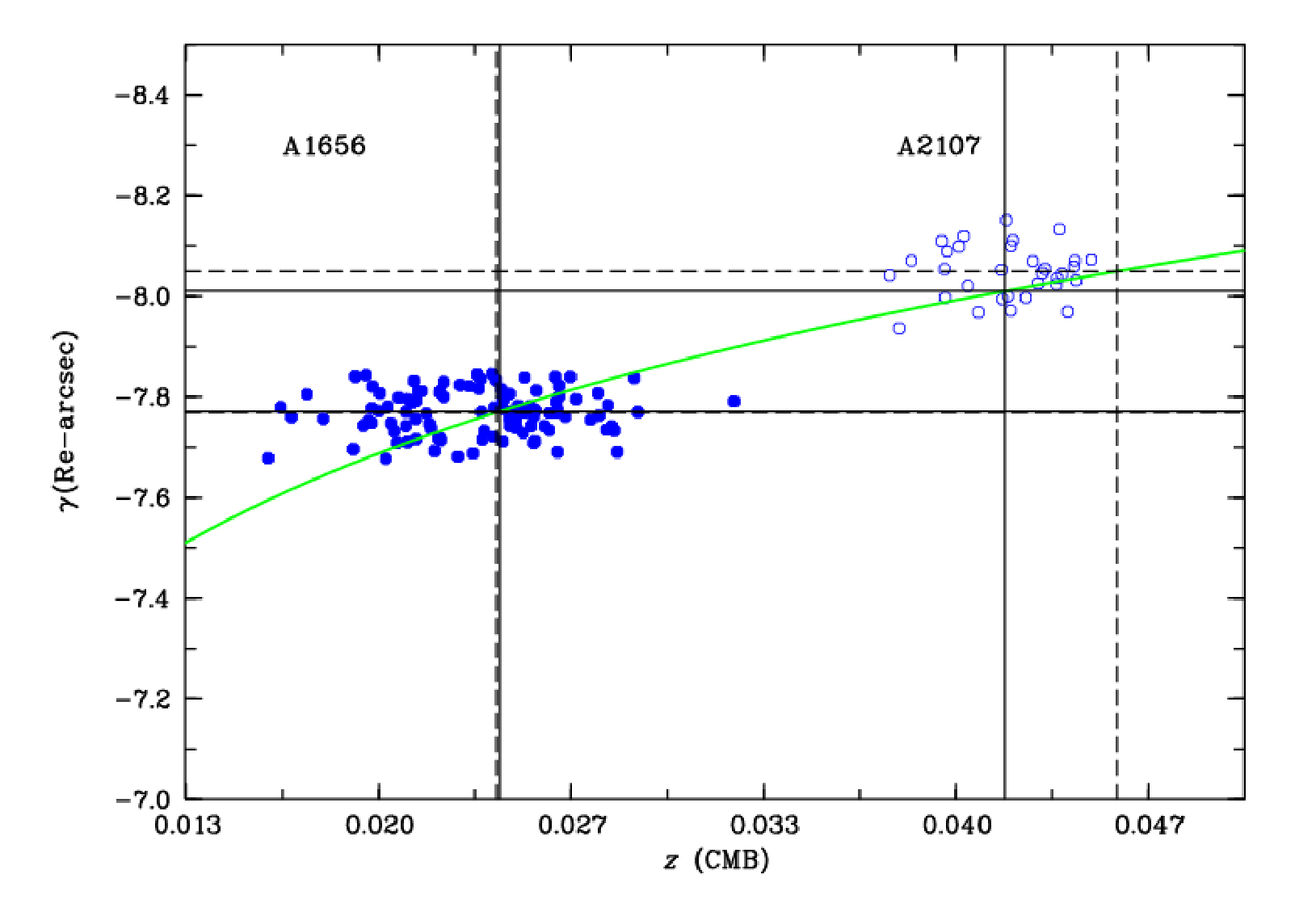}
\captionstyle{normal}
\caption{
The angular distances of galaxies, the FP zero points
$\gamma$, in dependence on the redshift $z_{CMB}$. THe galaxies are in
clusters A\,1656 (solid circles) and A\,2107 (open circles) within the
radius $R_{200}$. The bold curve corresponds to the Hubble relationship
between redshift and distance. Solid lines show the average redshifts
of clusters,$cz_{CMB}$, which yield the corresponding distances
at the ntersection with the Hubble curve. The dashed lines show the mean
distances of galaxy systems obtained by the FP and the  corresponding
redshifts, $z_{FP}$.
}
\label{A1656}
\end{figure*}

Figure.~\ref{A1656} graphically demonstrates the technique of
determining the peculiar velocities. In the diagram, there are the observed
distances (the zero points $\gamma$ calculated for $\log R_e$ with $R_e$
measured in arcseconds) for the galaxy clusters A\,1656 and A\,2107 (shown
with solid and open circles, respectively) in dependence on their redshifts
relative the CMB frame. The bold curve presents the expected Hubble
relationship between the distance and the redshift. It was calculated for
the model parameters we assumed:
$\Omega_m=0.3$, $H_0=70$~km~s$^{-1}$ Mpc$^{-1}$, $\Omega_{\Lambda}=0.7$
(equivalent to the parameter $q_o=-0.55$), and the FP zero point equal
to -8.066.
The angular distances of clusters of galaxies $\gamma$ are
converted to redshifts $z_{FP}$ by the  Pibbles approximation  \cite{Peebles}:
$$
D = \frac{cz}{H_o} \Bigl(1-\frac{(1-q_o)z}{2}\Bigr) = \frac{cz}{H_o}(1-0.225z),
$$
$$
D = \frac{cz}{H_o} \Bigl(\frac{1-0.225z)}{1+z}\Bigr).
$$
For the zero point of the expected Hubble dependence, we accounted for the
correction for the cosmological dimming effect in the SB of galaxies
$5\,\log\,(1+z_{\rm FP}$). The vertical solid lines show their average
redshifts relative to the CMB at each of the clusters, while
the horizontal solid lines, the corresponding  distances determined
relative to the expected Hubble dependence.
The dashed (horizontal) lines show the average distances of galaxy
clusters obtained by the FP, while the corresponding redshifts,
which were also determined relative to the expected Hubble dependence,
are shown by vertical dashed lines.

Figure~\ref{DiaH376} shows the Hubble diagram (upper panel)
that is the relative distances, i.e., the zero-points $\gamma$, in
dependence on the radial velocities (CMB). The clusters associated with
the void are presented by open blue circles. Thus, we may note that,
in Fig.~3,
the expected Hubble dependence almost correctly
describes the distances of groups/clusters of galaxies starting from the
Coma cluster ($z$ = 0.024) to the GV ($z\sim 0.15$).
In the lower panel, the deviations of the groups and clusters from the
Hubble flow are shown.

Figure~\ref{DiaH107} shows the same diagram, but with  accounting
for the evolution in the luminosity of early-type galaxies
$Q=1\,.\!\!^{\rm m}07z$.
It may be noted that the modeled Hubble dependence worse fits the data
when $cz$ < 15000~km/s and $cz$ > 30000~km/s.
The use of the other correction, namely that for the evolution
of the average surface brightness
$Q_r$ = 2.2z (expressed in $ mag/arcsec^2$) \cite{Saulder},
marginally improves the situation.

\section{DETERMINATION OF PECULIAR VELOCITIES BY THE FP}

The above Hubble dependence allows the
redshift of the cluster $z_{FP}$ to be determined from the corresponding
distance measured in arcseconds, $\log R_e$ (Fig.\ref{A1656}).
The peculiar velocities in the comoving coordinate system are defined
by  difference between the spectroscopical and photometrical redshifts, i.e.,
\begin{equation}
V_{pec} = c\,(z_{CMB}-z_{FP})/(1+z_{FP}),
\end{equation}
where $c$ is the velocity of light, $z_{CMB}$ is the redshift of the
cluster relative to the CMB, and $z_{FP}$ is the redshift of the cluster
corresponding to the distance determined by the FP.

\subsection{Comments about the cluster A\,1656 (Coma)}

The Coma cluster (A\,1656) has no peculiar velocity and is at
rest in the CMB system
(see, e.g. \cite{Hudson1, Colless, Gibbons, Tully1}).
This fact is often used to link the peculiar velocities of other the
galaxy clusters (e.g., \cite{Jorgensen, Mutabazi}).
In our study, we analyzed how the accounting for the evolution of
the luminosity influences the results on the Coma cluster.
In a paper \cite{Kopylova14}, we determined the relative distances of
clusters in superclusters by using the FP based on the SDSS data
\cite{Saulder}; for this, we used the evolution in luminosity of
early-type galaxies $Q=1\,.\!\!^{\rm m}07z$. If we take into account
this value, as well as the surface brightness dimming of galaxies
$10\,\log(1+z)$ and
and the restriction $M_r<-20\,.\!\!^{\rm m}6$, we will obtain
$V_{pec} = -388\pm120$~km/s.
If we consider all of the early-type galaxies, we will obtain
$V_{pec} = -840\pm120$~km/s. If we take into account the value
$Q_r$=2.2 (expressed in $mag/arcsec^2$) and the surface brightness
dimming of galaxies
$10\,\log(1+z)$, we will obtain $V_{pec} = -724\pm80$~km/s for the
Coma cluster. For the case we assumed, i.e. the average surface brightness
of early-type galaxies changes with $z$, $Q_r=3.76\pm0.01$ (expressed
in $mag/arcsec^2$), and the SB dimming follows $5\,\log(1+z)$, we obtain
the minimal peculiar velocity $V_{pec} = +40\pm70$~km/s for the Coma
cluster ($N$ = 107).

\subsection{Comments about the Virgo cluster}

The Virgo cluster is the nearest galaxy cluster ($z_h$ = 0.003821).
In paper \cite{Kopylov15}, we reported  the dynamical
parameters of the cluster  determined on the basis of the SDSS data.
In the present paper, for the region within 1.3$R_{200}$, we found only
eight early-type galaxies with the
parameters required for measuring the distance by the FP.
We considered these galaxies to measure the peculiar velocity of the
Virgo cluster and obtained $V_{pec}$ = -$240\pm260$~km/s. The derived
distance value was put on the Hubble diagram in Fig.~\ref{DiaH376}.
The similar peculiar velocity relative the observed distance of the
cluster is reported in a paper \cite{Tully2}.

As a result, we obtained that the peculiar velocity of the entire sample
of groups and clusters of galaxies ($N$ = 140) relative the CMB is
$V_{pec} = +190\pm90$~km/s. If only systems with the number of members
$N \geq 7$ are considered \cite{Jorgensen}, $V_{pec} = +170\pm90$~km/s
($N$ = 130).
In our sample, there are onle ten galaxy systems with $N$ < 7.
Moreover, our sample contains 106 galaxy systems, for which the radiation
was measured in the X-ray range \cite{Kopylova22}, and
34 systems with no data of this kind.
We found a week dependence of the measured relative distances $\gamma$
(and peculiar velocities)
on the X-ray luminosity in a band of 0.1--2.4 keV: the groups and clusters
with $L_X \leq 0.151 \times 10^{44}$~erg/s mainly exhibit positive
peculiar velocities (34 systems), while the clusters
with $L_X > 4 \times 10^{44}$~erg/s, the negative peculiar
velocities (A\,1795, A\,2142, and A\,2244).
For the sample in a interval $L_X = (0.151-4) \times 10^{44}$~erg/s,
we obtained $V_{pec} = -80\pm100$~km/s ($N$ = 63).
All of the galaxy systems, for which the X-ray radiation was
measured (N = 106) exhibit $V_{pec} = +160\pm90$~km/s.

These 34 galaxy systems are the groups/clusters, which are apparently
not virialized within the radius $R_{200}$, as is indicated by several
peaks in the radial velocity distribution (e.g., A\,1142, A\,1898,
and A\,2019),
or the groups similar to NGC\,5098. In the literature, we have found no
information about changes in the parameters of early-type galaxies and
their FP in the galaxy clusters in dependence on the X-ray radiation.
As for general changes in the parameters of galaxy clusters, we obtained
that the early-type galaxies (with $\log M_*$ = 10-11), in which the
formation of stars is suppressed (or does not exist at all), decrease
in size when they occur in the intergalactic medium of the galaxy cluster
\cite{Kopylova20}. It is clear that the larger the cluster's massâ
(i.e., the stronger the X-ray radiation), the stronger its influence on
the early-type galaxy. For the other parameters of galaxy systems---
such as the number of galaxies considered, the dispersion of the radial
velocities, the dynamical mass within the radius $R_{200}$,
and $z$ (though, for $z$ > 0.1, there are few early-type galaxies
in the galaxy systems, and the errors of determining their peculiar
velocities are large)---we detected no dependence of this kind.

With the quadratic error correction for the entire sample, the
root-mean-square deviation of the radial peculiar velocities is
$<V_{pec}^2>^{1/2}$ = $714\pm7$~km/s;
the sampling with $N \geq7$ yieids $<V_{pec}^2>^{1/2}$ = $740\pm7$~km/s.
For the sample in an interval of $L_X = (0.151-4) \times 10^{44}$~erg/s,
we derived $<V_{pec}^2>^{1/2}$ = $600\pm7$~km/s.

\subsection{Peculiar Motions of Galaxy Groups/Clusters Around
the Giant Void}

Figure~\ref{GV} presents  the distribution of groups and clusters around
the GV
on the diagram of the relative distances  (zero  points of systems with
$\log R_e$, where $R_e$ is measured in kiloparsecs) in dependence on the
radial velocity (CMB).
The solid line corresponds to the linear regression
$\gamma = 0.17(\pm0.29)z-8.08(\pm0.03)$ determined with using all
of the clusters ( N = 19), while the dashed lines show
the 1.5$\sigma$ deviations from it.

It may be noted that only one cluster A\,1609, falls these lines.
For this case, the linear regression is $\gamma = 0.05(\pm0.26)z-8.07(\pm0.03)$.
In this cases, the outflow velocities of galaxy clusters of galaxies
from the void are approximately
$250\pm410$~km/s and $70\pm370$~km/s, respectively.

The slope of the regression relationship obtained here is larger than that
we determined earlier with accounting for the evolution of luminosity of
galaxies \cite{Kopylov02}: 0.17 versus 0.033.
Consequently, the outflow velocities of galaxy clusters from the void are
approximately $250\pm410$~km/s and $47\pm447$~km/s,
respectively.

In other words, the results reported here and in \cite{Kopylov02}
do not contradict each other, though they were obtained by different
methods.
The main conclusions are the following: (1) we found no outflow of groups
and clusters of galaxies from the Giant Void;
(2) the peculiar motions of
the clusters around the GV are insignificant and do not exceed the
measurement errors, except for the cluster A\,1609, in which the ratio
of the peculiar velocity to the measurement error is 2.2.

\subsection{Peculiar Motions of Galaxy Superclusters}

In the region we considered ($z$<0.09), there are five large superclusters
of galaxies:
Hercules (Her, $z_h = 0.035$, $N$ = 11), Leo ($z_h = 0.036$, $N$ = 9),
Ursa Major (UMa, $z_h = 0.060$, $N$ = 11), Bootes (Boo, $z_h = 0.070$,
$N$ = 11), and Corona Borealis (CrB, $z_h = 0.072$, $N$ = 8).
The peculiar velocities of galaxy clusters within their boundaries  are
presented in our papers \cite{Kopylova07,Kopylova14,Kopylova17}.

We obtained the following peculiar velocities of the galaxy superclusters
themselves as averages for the constituent groups and clusters of galaxies
relative to the CMB: $V_{pec}$ = $+4\pm380$,
$+385\pm560$, $+467\pm660$, $+97\pm640$, and $+239\pm510$~km/s  for
the the Her, Leo, UMa, Boo, and CrB superclusters, respectively.
The average peculiar velocity for all of the galaxy superclusters
is +$240\pm250$~km/s. A small excess of the positive
peculiar velocities is connected with a large number of galaxy groups with
$L_X \leq 0.151 \times 10^{44}$~erg/s in superclusters (Section 4).

Recently, in a paper \cite{Kopylova24}, we built for the first time
the FPs of the galaxy groups and clusters themselves analogously to
that of elliptical galaxies. We showed that their distances correspond
to the expected Hubble dependence, though their dispersion from the
latter is three times larger than that obtained here by elliptical
galaxies. We also measured the average peculiar velocity for all
superclusters of galaxies, which turned out to be
+$75\pm360$~km/s.

\section{THE HUBBLE DIAGRAM AND DEVIATIONS FROM IT}

There is a contradiction in the determining the Hubble constant $H_o$,
one of the fundamental cosmological parameters. The constant $H_o$
estimated by the Cepheid-supernova distance ladder differs from the
value extrapolated from the CMB data under the assumption of a
standard cosmological model: $74.0\pm1.4$ km~$s^{-1}$ Mpc$^{-1}$ \cite{Riess}
versus
$67.4\pm0.5$ km~$s^{-1}$ Mpc$^{-1}$ \cite{Planck Collaboration},
respectively.

\begin{figure*}[p]
\setcaptionmargin{5mm}
\onelinecaptionsfalse
\includegraphics[scale=0.63,angle=0]{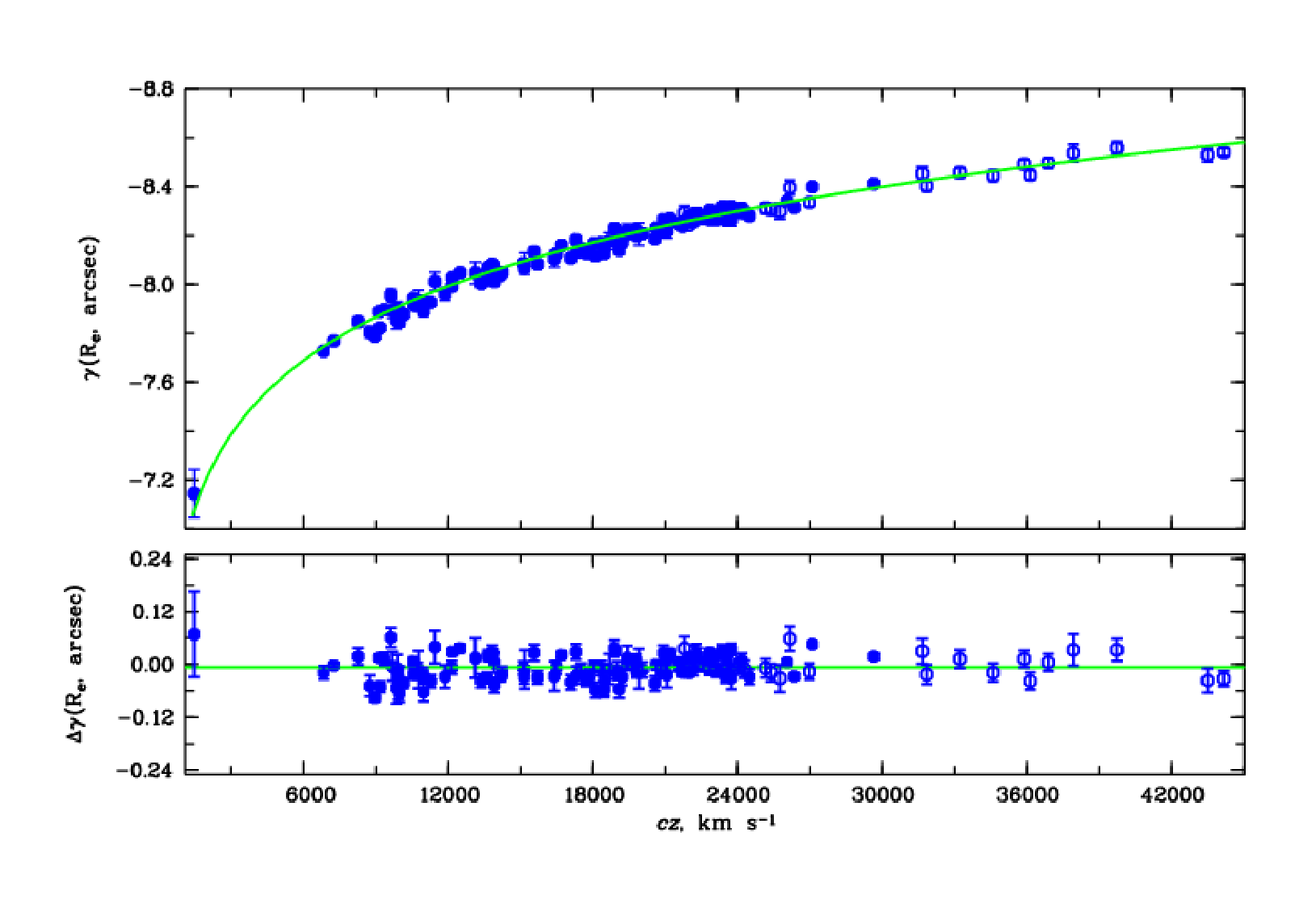}
\captionstyle{normal}
\caption{
Upper: The angular distances of 140 groups/clusters of galaxies, the FP
zero points $\gamma$, in dependent on the radial velocity $cz$
(the Hubble diagram); the distances were obtained with accounting for the
evolutionary parameter $Q_r=3.76$~(expressed in $mag/arcsec^2$).
The open circles are for the systems located around the GV ($N$ = 19).
The solid line shows the expected Hubble dependence in the cosmological
$\Lambda$CDM model with $\Omega_m$ = 0.30.
Lower panel: the curve of the residual deviations.
}
\label{DiaH376}
\end{figure*}

\begin{figure*}[p]
\setcaptionmargin{5mm}
\onelinecaptionsfalse
\includegraphics[scale=0.63,angle=0]{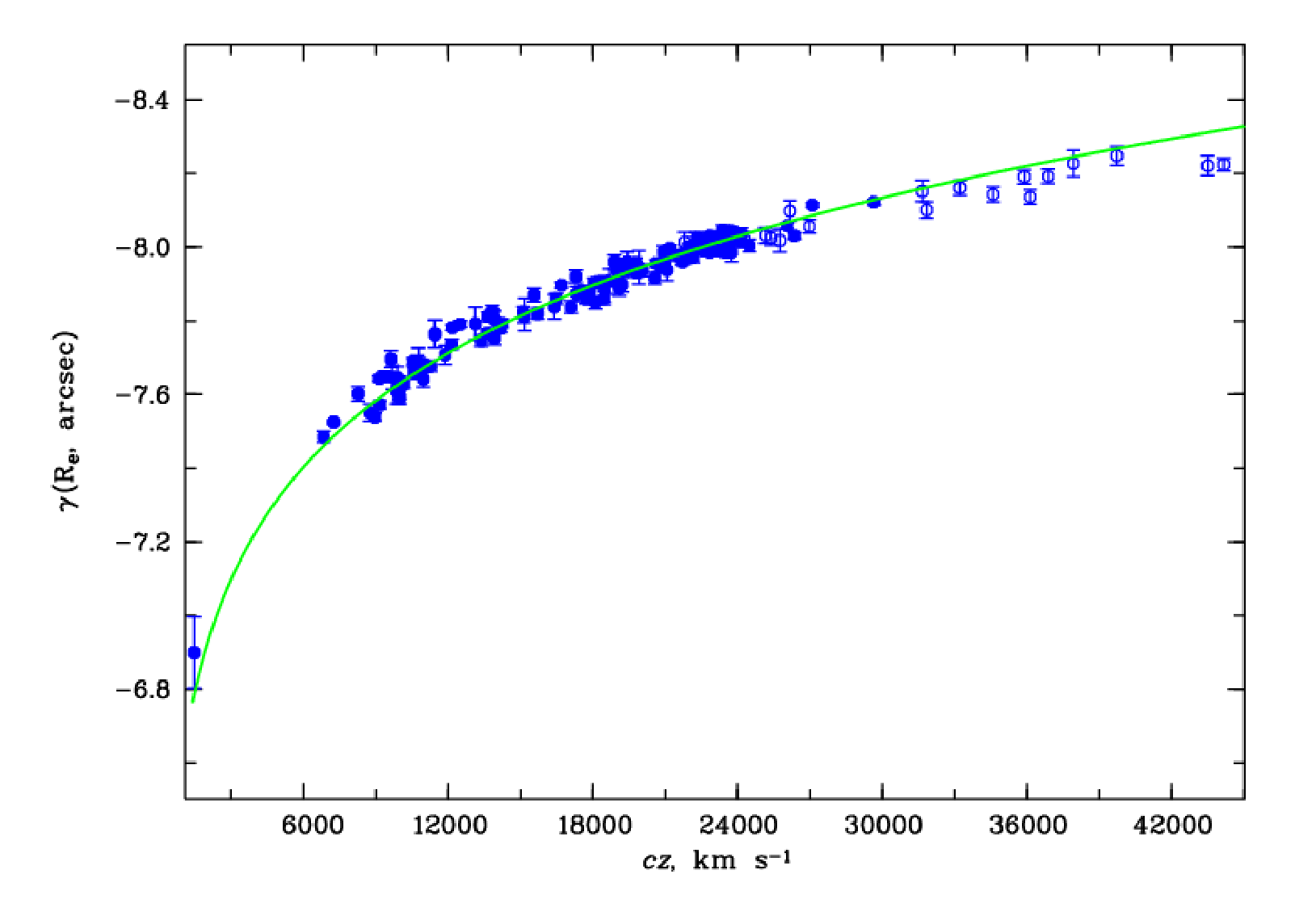}
\captionstyle{normal}
\caption{
The angular distances of 140 groups/clusters of galaxies, the FP zero points
$\gamma$, in dependence on the radial velocity $cz$;
the distances were obtained with accounting for the evolutionary parameter
$Q = 1\,.\!\!^{\rm m}07z$.
Designations are the same as those in Fig.~3.
}
\label{DiaH107}
\end{figure*}

\begin{figure*}[p]
\setcaptionmargin{5mm}
\onelinecaptionsfalse
\includegraphics[scale=0.63,angle=0]{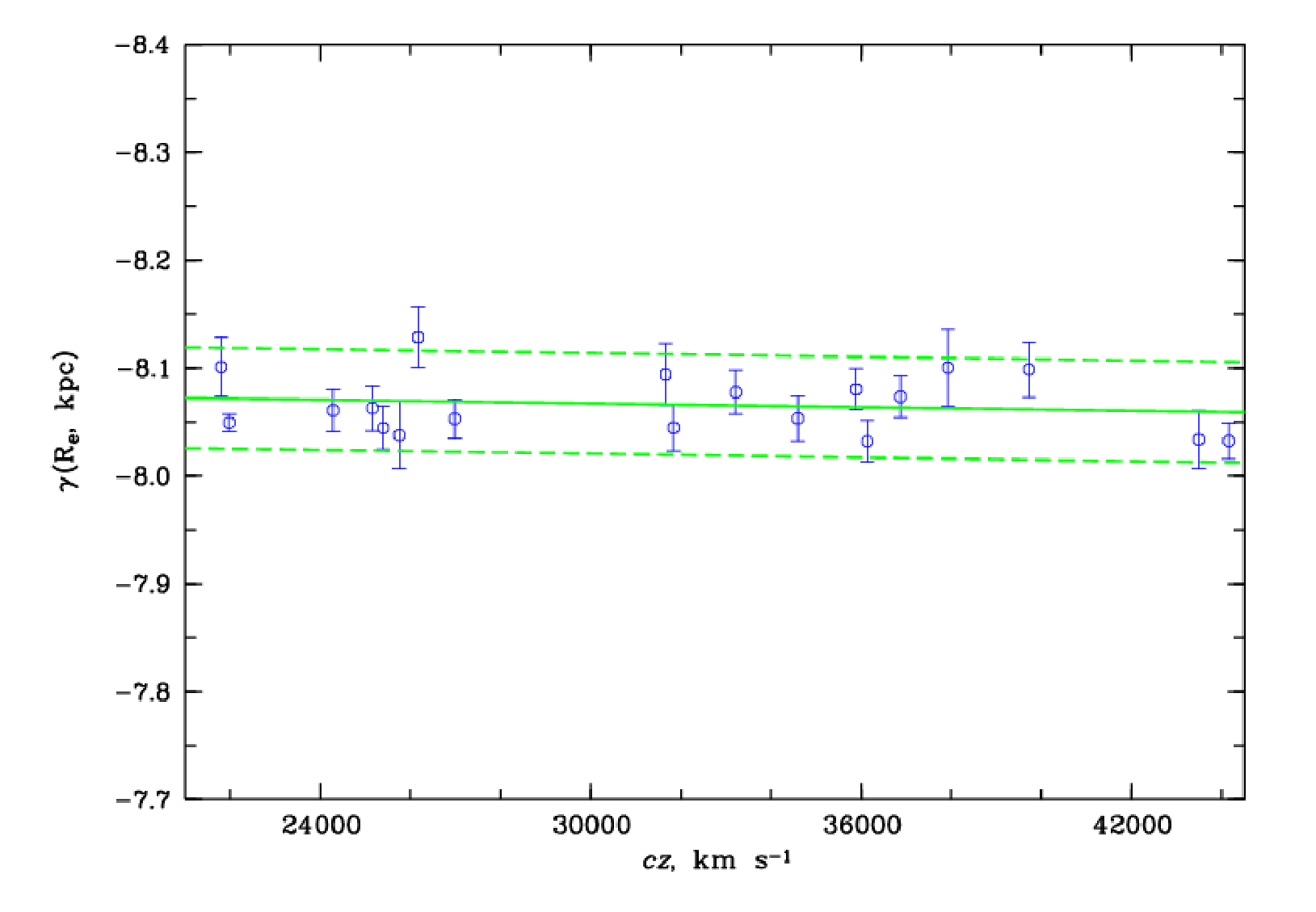}
\captionstyle{normal}
\caption{
Individual distances of groups/clusters of galaxies
around the GV ($\log R_e$, where $R_e$ is  measured in kiloparsecs) in
dependence on the radial velocity $cz$.
The solid line presents the linear regression
$\gamma = 0.17(\pm0.29)z-8.08(\pm0.031)$ that was determined
with the use of all of the clusters ($N$ = 19), while
the dashed lines show the deviations from it at a level 1.5$\sigma$.
}
\label{GV}
\end{figure*}

In Fig.~\ref{DiaH376}, the solid green curve represents the Hubble
relationship between the radial velocity in the CMB system and the
angular distance.
The curve corresponds to the flat $\Lambda$CDM model with
$\Omega_{\Lambda} = 0.7$,
$\Omega_m=0.3$, and the Hubble constant $H_0=70$~km~$s^{-1}$ Mpc$^{-1}$.

In the lower panel of Fig.~\ref{DiaH376}, we  show the deviations from the
Hubble dependence for the distances we obtained for the galaxy systems.
We derived the mean deviation from the Hubble dependence
$<\Delta \gamma>$ = -$0.0066\pm0.0023$ and
($N$ = 130) $<\Delta \gamma>$ = -$0.0065\pm0.0023$ for the entire
sample ($N$ = 140) and the galaxy systems with more than seven members
($N$ = 130), respectively. The corresponding standard deviations are
0.0275 and 0.0264, which are related to deviations of 6.3\% and 6.08\%
($\pm4.4$ and $\pm4.2$~km~s$^{-1}$ Mpc$^{-1}$), respectively) in the
Hubble constant. The mean positive and negative deviations are approximately
the same and correspond to $<\Delta \gamma>$ = +$0.0218\pm0.0020$ and
$<\Delta \gamma>$ = -$0.0244\pm0.0018$  with $N$ = 54 and 86, respectively.
For the sample in an interval $L_X = (0.151-4) \times 10^{44}$~erg/s,
we obtained the mean deviation from the Hubble dependence
$<\Delta \gamma>$ = $0.0017\pm0.0028 $ ($N$ = 63) with a standard scatter
of 0.0224 corresponding to the deviation of
5.11\% ($\pm3.6$~km~s$^{-1}$ Mpc$^{-1}$).

If we deduct the peculiar velocities of groups/clusters of galaxies
(the formula (1)) from the data in Fig~\ref{DiaH376}, the errors in
measuring the distances by the FP will actually determine the deviations
on the Hubble diagram. In this case, the standard deviations are
0.0173 ($N$ = 140) and
0.0163 ($N$ = 140) for the galaxy systems with more than seven members,
which correspond to deviations of $\pm2.8$ and $\pm2.6$~km~$s^{-1}$ Mpc$^{-1}$,
respectively, in the Hubble constant. For the sample in an
interval $L_X = (0.151-4) \times 10^{44}$~erg/s, we obtained
a corresponding standard deviation of 0.0130 for ($N$ = 63),
which is related to a deviation of $\pm2.1$~km~$s^{-1}$ Mpc$^{-1}$) in
the Hubble constant.

\section{CONCLUSIONS}

To understand the origin and evolution of a large-scale structure of the
Universe, it is important to study the peculiar motions of groups and
clusters of galaxies both in massive superclusters of galaxies and around
voids, which should expand faster than the Hubble flow.
Model calculations show that high peculiar velocities of galaxy systems,
$V_{pec}>10^3$~km/s, occur in dense superclusters of galaxies
\cite{Bahcall}.

In order to study the peculiar velocities of galaxy systems, we selected a
sample of groups/clusters of galaxies from large superclusters---
Hercules, Leo, Ursa Major, Corona Borealis, and Bootes---and from smaller
galaxy systems
\cite{Kopylova07,Kopylova14,Kopylova17,Kopylova21} as well as a sample of
systems of galaxies around the GV \cite{Kopylov02}. With the use of the
FP of early-type galaxies, we determined the relative distances of galaxy
systems and measured their peculiar velocities.

We obtained that the peculiar velocities vary from $\pm10$ to
$\pm$3000~km/s. Twelve groups/clusters of galaxies have peculiar
velocities exceeding measurement errors by more than three times.
The average peculiar velocity relative to the CMB frame for the sample of
galaxy clusters ($N$ = 130), each of which contains more than seven
members, is $+172\pm90$~km/s.
The root-mean-square deviation of the radial peculiar velocities with
quadratic accounting for errors is
$<V_{pec}^2>^{1/2}$ = $740\pm7$~km/s.
The average peculiar velocity of five galaxy superclusters of galaxies is
$+240\pm250$~km/s.

We found a weak dependence of distances and peculiar velocities on the X-ray
luminosity of galaxy groups and clusters with
$L_X  \leq 151 \times 10^{44}$~erg/s.
In an interval of  $L_X = (0.151-4) \times 10^{44}$~erg/s ($N$ = 63),
almost  no dependence of this kind is observed; and the average peculiar
velocity obtained for this sample is $-80\pm100$~km/s, while the standard
deviation of peculiar velocities with quadratic accounting for errors
is $<V_{pec}^2>^{1 / 2}$ = $600\pm7$~km/s.

In this study, we verified whether a cosmological test such as the Hubble
diagram in the standard $\Lambda$CDM model is consistent with the data
of observations. We measured the mean deviation from the Hubble dependence
for galaxy groups and clusters with and without accounting for the peculiar
velocities. The deviation from the Hubble dependence (minus peculiar velocities
of galaxy systems) is determined by errors in measuring the distances.
For this case, we obtained a value of $\pm0.0173$ ($N$ = 140) for the standard
logarithmic deviation and the corresponding standard deviation for the
Hubble constant $H_0 = 70\pm2.8$~km~s$^{-1}$ Mpc$^{-1}$. For the
sample with $L_X = (0.151-4) \times 10^{44}$~erg/s, we obtained
a standard deviation of $\pm0.0130$ ($N$ = 63), which corresponds to the
deviation in the Hubble constant $H_0 = 70\pm2.1$~km~s$^{-1}$ Mpc$^{-1}$.

As in the previous paper \cite{Kopylov02}, we found no substantial
outflow of galaxy  groups and clusters from
the GV: the outflow velocity measured by 19 galaxy systems is approximately
$250\pm410$~km/s. The peculiar motions of galaxy systems around GV
are insignificant and do not exceed the measurement errors, except for the
cluster A\,1609, the ratio of the peculiar velocity of of which
to the measurement error is 2.2.

\begin{acknowledgments}
This research has made use of the NASA/IPAC Extragalactic Database
(NED, \url{http://nedwww.}\linebreak\url{.ipac.caltech.edu}),
which is operated by the Jet Propulsion Laboratory, California Institute of
Technology, under contract with the National Aeronautics and Space
Administration, Sloan Digital Sky Survey (SDSS, \url{http://www.sdss.org}),
which is supported by Alfred P. Sloan Foundation, the participant institutes
of the SDSS collaboration, National Science Foundation, and the United
States Department of Energy and Two Micron All Sky Survey (2MASS,
\url{http://www.ipac.}\linebreak\url{.caltech.edu/2mass/releases/allsky/}).
\end{acknowledgments}

\begin{center}
\refname
\end{center}

Translated by E. Petrova

\newpage

\newpage

\newpage

\newpage

\newpage


\begin{thebibliography}{45}
\bibitem{Gregory}
S.~A.~Gregory, L.~A.~Thompson, \apj {\bf 222}, 784 (1978).
\bibitem{Joeveer}
J{\~o}eveer, M., Einasto, J., Tago, E. \mnras, {\bf 185}, 357 (1978).
\bibitem{Kirshner}
Kirshner, R.~P., Oemler, A.~Jr., Schechter, P.~L. et al. \apj, {\bf 248},
L57 (1981).
\bibitem{de Lapparent}
de Lapparent, V., Geller, M.~J., Huchra, J.~P. \apj, {\bf 302}, L1 (1986).
\bibitem{Kopylov88}
Kopylov, A.~I., Kuznetsov, D.~Y., Fetisova, T.~S., Shvartsman, V.~F. in
In Large Scale Structures of the Universe, ed. Audouse J. et al., IAUS, 130,
129 (1988).
\bibitem{Bond}
Bond, J.~R., Kofman, L., Pogosyan, D. \nat, {\bf 380}, 603 (1996).
\bibitem{Einasto01}
Einasto, M., Einasto, J., Tago, E., M{\"u}ller, V., Andernach, H. \aj,
{\bf 122}, 2222 (2001).
\bibitem{Zeldovich}
Zeldovich, Ia.~B., Einasto, J., Shandarin, S.F. \nat, {\bf 300}, 407 (1982).
\bibitem{Batuski}
Batuski, D.~J. and Burns, J.~O. \aj,  {\bf 90}, 1413 (1985).
\bibitem{Tully}
Tully, R.~B. \apj, {\bf 303}, 25 (1986).
\bibitem{Stavrev}
Stavrev, K.~Y. \aas, {\bf 144}, 323 (2000).
\bibitem{Kopylov02}
Kopylov, A.~I. and Kopylova, F.~G. \aaa, {\bf 382}, 389 (2002).
\bibitem{Einasto11}
Einasto, J., Suhhonenko, I., H{\"u}tsi, G. et al. \aaa, {\bf 534A}, 128
(2011).
\bibitem{Dressler}
Dressler, A., Lynden-Bell, D., Burstein, D. et al. \apj, {\bf 313}, 42
(1987).
\bibitem{Djorgovski}
Djorgovski, S., Davis, M. \apj, {\bf 313}, 59 (1987).
\bibitem{Wegner}
Wegner, G., Colless, M., Baggley, G. et al. \apjs,  {\bf 106}, 1 (1996).
\bibitem{Hudson}
Hudson, M.~J., Smith, R.~J., Lucey, J.~R. et al. \apj, {\bf 512}, L79 (1999).
\bibitem{da Costa}
da Costa, L.~N., Bernardi, M., Alonso, M.~V. et al. \apj, {\bf 537}, L81
(2000).
\bibitem{Batiste}
Batiste, M. \&  Batuski, D.~J. \mnras, {\bf 436}, 3331 (2013).
\bibitem{Aihara}
Aihara, H., Allende Prieto, C., An, D. et al. \apjs, {\bf 193}, 29 (2011).
\bibitem{Kopylova14}
Kopylova, F.~G., Kopylov, A.~I. Astron. Lett., {\bf 40}, 595 (2014).
\bibitem{Kopylova17}
Kopylova, F.~G., Kopylov, A.~I. \bsao, {\bf 72}, 363 (2017).
\bibitem{Kopylova21}
Kopylova, F.~G., Kopylov, A.~I. Astron. Astrophys. Tran., {\bf 32}, 105
(2021).
\bibitem{Kopylova07}
Kopylova, F.~G., Kopylov, A.~I. Astron. Lett., {\bf 33}, 211 (2007).
\bibitem{Saulder}
Saulder, C., Mieske, S., Zeilinger, W.~W., Chilingarian, I. \aaa,
{\bf 557A}, 21 (2013).
\bibitem{Mohr}
Mohr, J.~J., Wegner, G. \aj, {\bf 114}, 1 (1997).
\bibitem{Carlberg}
R.G.~Carlberg, H.K.C.~Yee, E.~Ellingson et al., \apj \textbf{485}, L13
(1997).
\bibitem{Kopylova13}
Kopylova, F.~G. \bsao, {\bf 68}, 253 (2013).
\bibitem{Strauss}
M.A.~Strauss, D.H.~Weinberg, R.H.~Lupton et al., Astron. J. {\bf 124}, (2002).
\bibitem{Chilingarian}
I.V.~Chilingarian, A.~Melchior, and I.Y.~Zolotukhin, \mnras {\bf 405},
1409 (2010).
\bibitem{Jorgensen}
I.~Jorgensen, M.~Franx, and P.~Kjaergaard, \mnras \textbf{280}, 167
(1996).
\bibitem{Peebles}
P.~J.~E.~Peebles, In ``Principles of Physical Cosmology'', 1993.
\bibitem{Hudson1}
M.J.~Hudson, J.~R.~Lucey, R.~J.~Smith, and J.~Steel, \mnras {\bf 291},
488 (1997).
\bibitem{Colless}
M.~Colless, R.~P.~Saglia, D.~Burstein et al., \mnras \textbf{321}, 277
(2001).
\bibitem{Gibbons}
R.~A.~Gibbons, A.~S.~Fruchter, and G.~D.~Bothun, \aj \textbf{121}, 649
(2001).
\bibitem{Tully1}
R.~B.~Tully, H.~M.~Courtois, A.~E.~Dolphin et al., \aj {\bf 146}, 86 (2013).
\bibitem{Mutabazi}
T.~Mutabazi, \apj, \textbf{911}, 2102 (2021).
\bibitem{Kopylov15}
Kopylov, A.~I., Kopylova, F.~G. \bsao, {\bf 70}, 243 (2015).
\bibitem{Tully2}
E. J. Shaya, R.~B.~Tully, D.~Pomar\'{e}de and A.~Peel, \apj {\bf 927}, 168 (2022).
\bibitem{Kopylova22}
Kopylova, F.~G., Kopylov, A.~I. \bsao, {\bf 77}, 347 (2022).
\bibitem{Kopylova20}
Kopylova, F.~G., Kopylov, A.~I. \bsao, {\bf 75}, 424 (2020).
\bibitem{Kopylova24}
Kopylova, F.~G., Kopylov, A.~I. \bsao, {\bf 79}, 1 (2024).
\bibitem{Riess}
A.~G.~Riess, S.~Casertano, W.~Yuan et al., \apj, \textbf{876}, 85 (2019).
\bibitem{Planck Collaboration}
Planck Collaboration c., \aaa, \textbf{641}, A6 (2020).
\bibitem{Bahcall}
N.~A.~Bahcall, M.~Gramann, R.~Cen, \apj, \textbf{436}, 23 (1994).
\end{thebibliography}
\end{document}